\documentclass{acm_proc_article-sp}

\toappear{
}

\usepackage[utf8]{inputenc}

\usepackage{balance}  
\usepackage{graphics} 
\usepackage{times}    
\usepackage{url}      

\makeatletter
\def\url@leostyle{%
  \@ifundefined{selectfont}{\def\UrlFont{\sf}}{\def\UrlFont{\small\bf\ttfamily}}}
\makeatother
\begin{document}

\conferenceinfo{WWW}{'15 Florence, Italy}

\title{Investigating Rumor Propagation with TwitterTrails}

\numberofauthors{3}
\author{
  \alignauthor Samantha Finn \\
    \affaddr{Computer Science}\\
    \affaddr{Wellesley College}\\
    \email{sfinn@wellesley.edu}
  \alignauthor Panagiotis Takis Metaxas\thanks{Corresponding author.} \\
    \affaddr{Computer Science}\\
    \affaddr{Wellesley College}\\
    \email{pmetaxas@wellesley.edu}\\
%
  \alignauthor Eni Mustafaraj\\
    \affaddr{Computer Science}\\
    \affaddr{Wellesley College}\\
    \email{emustafa@wellesley.edu}
}

\maketitle
\begin{abstract}
Social media have become part of modern news reporting, used by journalists to spread information and find sources, or as a news source by individuals. The quest for prominence and recognition on social media sites like Twitter can sometimes eclipse accuracy and lead to the spread of false information. As a way to study and react to this trend, we introduce {\sc TwitterTrails}, an interactive, web-based tool ({\tt twittertrails.com}) that allows users to investigate the origin and propagation characteristics of a rumor and its refutation, if any, on Twitter. 
Visualizations of burst activity, propagation timeline, retweet and co-retweeted networks help its users trace the spread of a story. Within minutes {\sc TwitterTrails} will collect relevant tweets and  automatically answer several important questions regarding a rumor: its originator, burst characteristics, propagators and main actors according to the audience. In addition, it will compute and report the rumor's level of visibility and, as an example of the power of crowdsourcing, the audience's skepticism towards it which correlates with the rumor's credibility. 
We envision {\sc TwitterTrails} as valuable tool for individual use, but we especially for amateur and professional journalists investigating recent and breaking stories. Further, its expanding collection of investigated rumors can be used to answer questions regarding the amount and success of misinformation on Twitter. 
\end{abstract}

\section{Introduction} 
The so-called ``24 hour news cycle'' has led to an increased sensationalism of news stories.  Especially with the increase in cable news channels and online news media, the need to catch the attention of the public has led to faster and more hyped up reporting. Many compete to be the first to report a breaking story and present new and exclusive angles.\footnote{{\em 24 Hour News Killed Journalism}, by Jeff Sorensen. Huffington Post Blog, Aug.20, 2012 http://huff.to/1tRlttE}
This trend has fed off social media and in turn empowered citizen journalists publishing and transmitting news through websites like Twitter and Facebook. Most of the time the information is true, but the desire to be first and receive more likes and retweets sometimes trumps accuracy and fact checking. It many cases, it may not matter much whether a rumor is true or false, but there are some cases that it matters greatly. 

Consider the following scenario, that will serve as a running example in our description: Around noon on March 27, a reporter sees a tweet indicating that an airplane was spotted in the sea near the Canary Islands. For context, this happens just a few weeks after the disappearance of the Malaysian Airlines 370 flight on March 8, which captured the attention of people world wide.

\begin{figure}[t]
\centering
\includegraphics[width=.28\textwidth,keepaspectratio]{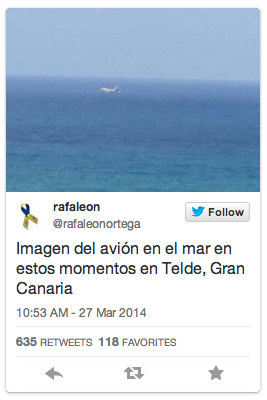}
\caption{A tweet spreading around 12 noon EST on March 27, 2014, reads (in Spanish) ``Picture of the airplane in the sea these moments in Telde, Grand Canary Island''.}
\label{fig:plane-tweet}
\end{figure}

Pressing the retweet button is very tempting in this situation, but spreading this information further should not be done automatically. It would be very helpful if the reporter can quickly determine a few facts about this story\footnote{In this paper we use the term story to indicate a rumor, true or false, spreading through Twitter.}, including:

\begin{itemize}
\item {\bf Originator}: Who ``broke'' the story first (made it widely known)?
\item {\bf Burst}: When and how did the story break (that is, have the first burst in its propagation)?
\item {\bf Timeline}: How is the story propagating over time? Is it still spreading at the time of the investigation? 
\item {\bf Propagators}: Who has been spreading the story by retweeting,
given that \cite{RT-agreement} retweets often indicate agreement with the message?  
\item {\bf Negation}: Were there any refutations of the story competing for attention? How widespread were they, compared to the original claim?
\item {\bf Main actors}: Who were the main actors in the propagation, according to the Twitter audience?
\end{itemize}

There is no formal quality control in the realm of citizen reporting.  Reliable information can be created by witnesses and spread through social media networks, which could aid journalists when writing a story.  But how can journalists or other individuals verify the claims of information they discovered on Twitter?  Searching the Internet and social media can be tedious and time consuming, and might require technical information that an individual doesn't have readily available.  
In the case of trending stories, massive amounts of data are being created and circulated, and often there will be individuals or bots trying to manipulate this data to promote their agenda \cite{WebSpamPropaganda}.

In this paper, we present {\sc TwitterTrails}, a new web-based tool for interactive exploration of Twitter information, which helps answer the above-listed questions.  
The major features of the {\sc TwitterTrails} system are summarized below:

{\sc TwitterTrails} 
retrieves relevant data from Twitter based on inputs from the tweet which the user is investigating, and allows the user to modify and refine this dataset.  From this set of relevant tweets, it provides a summary of its findings along with several interactive visualizations to allow the user to explore and analyze the data: 
the Propagation Graph, which highlights the tweets and accounts which broke the story on Twitter; the Timeline Graph which shows the activity of the story over time and allows the user to selectively browse the data; and the Retweet and Co-Retweeted Network Graphs, which highlight accounts that were influential in spreading the story.
The collection of rumors in the {\sc TwitterTrails} system
can be used to answer questions regarding the amount and success of misinformation on Twitter. 

We designed the system specifically for Twitter because it is easy through their APIs to collect the data. Given appropriate APIs one can build such a system for Facebook and other social networks. Unfortunately, such APIs are not available at the time of writing.

\begin{figure}[ht!]
\centering
\includegraphics[width=.50\textwidth,keepaspectratio]{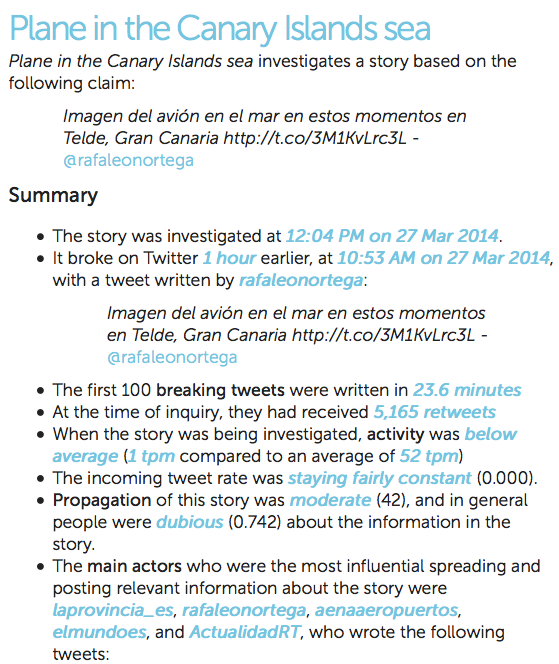}
\caption{The automatically generated summary provides immediate feedback to the user investigating the rumor.}
\label{fig:summary}
\end{figure}

\section{Overview of  TwitterTrails}
Credibility of information is strongly related to trust in the source. Before a savvy Twitter user retweets a tweet, she should feel reasonably confident in the validity of the information presented in that tweet, or else she might risk 
damaging her own reputation.  This is true for aspiring citizen journalists, and even more important for professional journalists on social media. 
{\sc TwitterTrails} 
is an investigative and exploratory tool, to analyse the origin and spread of a story on Twitter. While it does not answer directly the question of a story's validity, it provides information that a critically thinking person can use to examine how a Twitter audience reacts to the spreading of the story. We currently envision {\sc TwitterTrails}  as a tool for journalists utilizing Twitter as a source of information, but in the future we want it to be useful to Twitter users with a working understanding of our visualizations. 

{\sc TwitterTrails}  takes as an input from the user a single tweet with information she wishes to investigate, like the one in Figure~\ref{fig:plane-tweet}, 
but allows the user to input keywords from that tweet to collect a set of related tweets.  From that set of related tweets it provides visualizations to pinpoint the origin of the investigative tweet: where the information trail started, who initially broke the story.  In some cases this may be enough for the user, based on the reputation of the accounts which broke the story on Twitter by weighing factors such as whether they are verified, if they have many followers, the age of the account, or studying their profile and recent tweets.

In cases of more dubious data, or for a more engaged Twitter user or journalist, {\sc TwitterTrails}  provides visualizations to trace not only the origin, but the spread of a story.  It gives the user tools to answer important questions about the story: who wrote its first tweet, and who popularized it?  When did the story break and how did it spread?  When was it most active, and what information and users were prominent at peak times?  What users were influential in the spread of the story, and who did users put their trust in when spreading the story? 
Propagation and Timeline visualizations give the user a meaningful way to browse the data, while network graphs give her an overview of influential users in the data. Moreover, minutes after the launching of an investigation {\sc TwitterTrails} will give the user a summary of the findings that in most cases may be enough to answer her questions (Fig.~\ref{fig:summary}). If she wants details on how this summary is produced, she can look into each of the sections that the investigation produced.

We first give an overview of the architecture and usage of the tool and then we discuss the main algorithms and metrics that power {\sc TwitterTrails}.  We do so by referring to a specific story studied, but the interested reader should examine the live system at {\tt twittertrails.com}.

\begin{figure*}[ht!]
\centering
\includegraphics[width=6in,keepaspectratio]{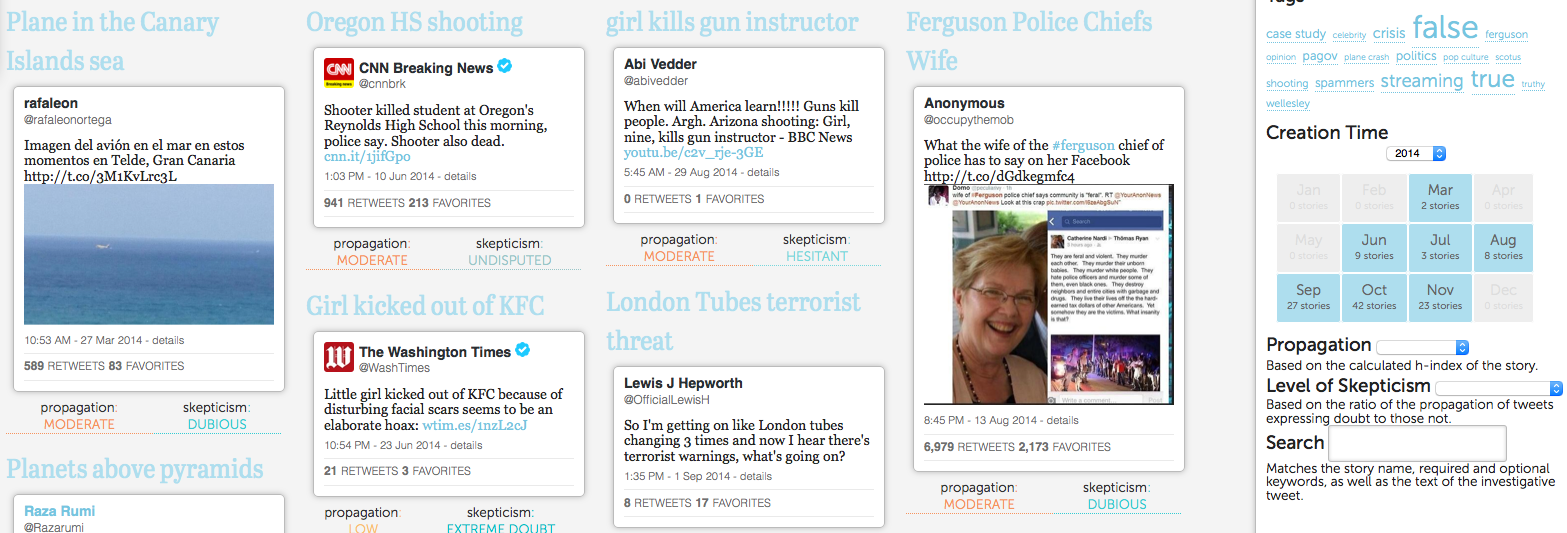}
\includegraphics[width=6in,keepaspectratio]{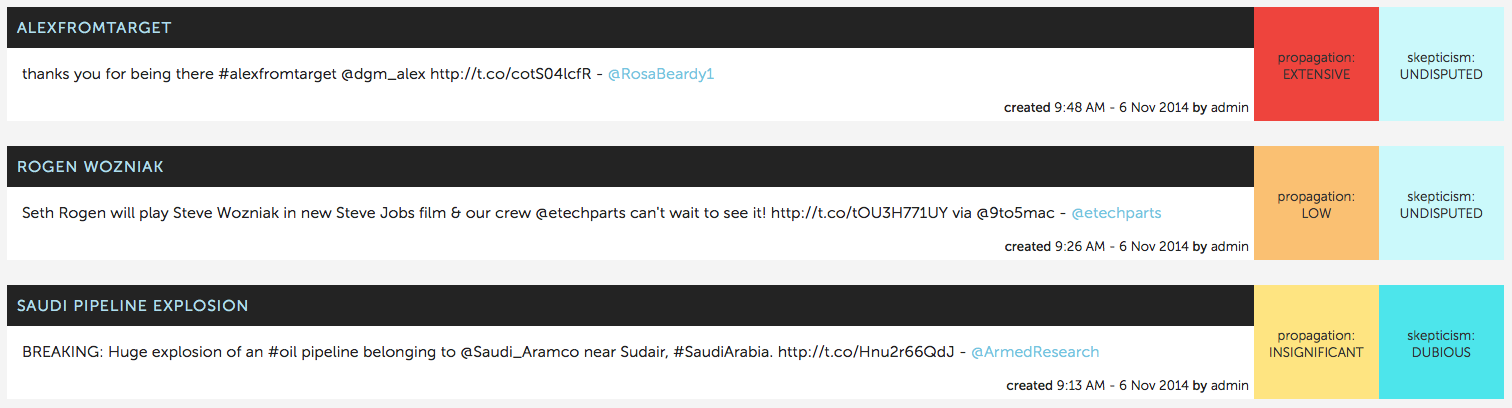}
\caption{The full story  view page (top) and condensed view page (bottom) of {\tt twittertrails.com}.}
\label{fig:twittertrails-com}
\end{figure*}

\section{Architecture of the System}

\subsection{ Investigative Tweet }

{\sc TwitterTrails}  is structured around the investigation of a single tweet, which is the first input the user provides (via the url of the tweet).  Throughout this paper we will reference the Plane in the Sea story, which was investigated from a tweet by @rafaleonortega on March 27th, 2014, reporting that there was a plane in the sea off the coast of Telde in the Canary Islands (see Figure ~\ref{fig:plane-tweet}) only a few weeks after the disappearance of Malaysian Airlines Flight 370.  
This tweet includes a picture of what looks like a plane in the water.   

After retrieving the investigative tweet, {\sc TwitterTrails} provides the Keyword Selection interface, to allow the user to highlight words and phrases from the tweet as keywords, or enter them manually. The system helps the user select the appropriate keywords in a variety of ways.  
to assist the user in chosing terms which will produce the most relevant data for their story, the system gives a rating to potential search terms.  It fetches the last 100 tweets containing the term, and scores how related these tweets are to each other and to the investigative tweet. 
It also suggests new search terms from common words, bigrams and hashtags in the 100 most recent tweets. 
For @rafaleonortega's tweet, we chose the following keywords: "gran canaria", "imagen", "airplane", and "telde".

\subsection {Refining Relevant Tweets}
The algorithm for searching for, collecting and determining relevant tweets is described in section~\ref{relevancy}.  
{\sc TwitterTrails } provides an interface for the user to modify the inputs which go into the determining the relevant tweets as many times as she likes in order to select the best set of relevant tweets. 
For the Plane in the Sea story, we added "avión" as the only required keyword, with none of the other search terms needing to be present.

\subsection{Story Interface}

\begin{figure*}[ht!]
\centering
\includegraphics[width=6in,keepaspectratio]{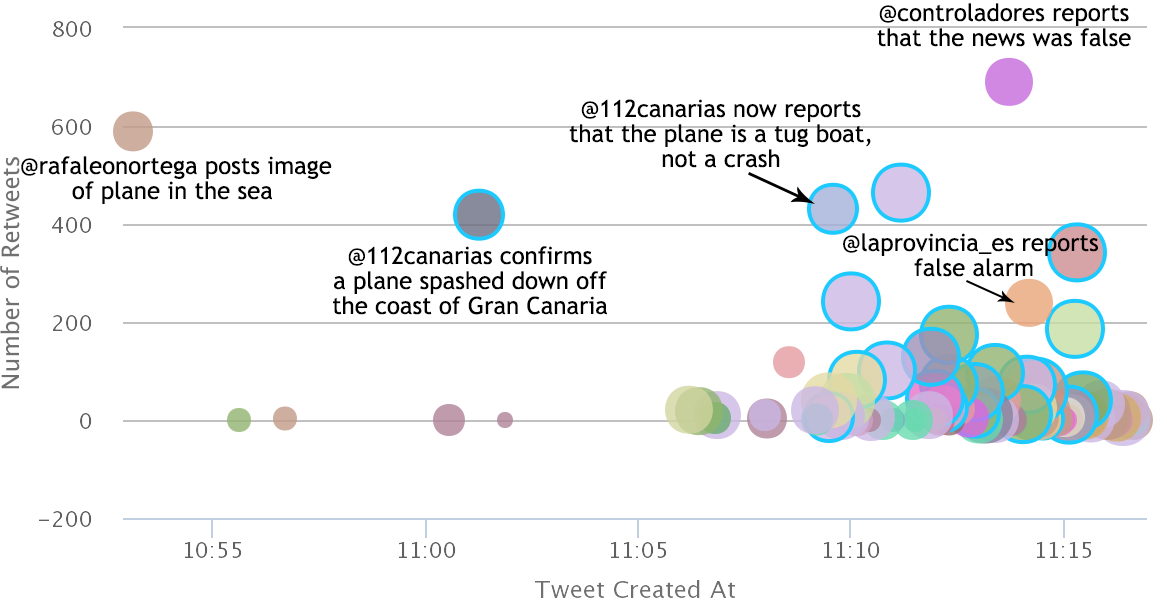}
\caption{The Propagation Graph from the Plane in the Sea story.}
\label{fig:propagation}
\end{figure*}

\subsubsection{Tweet Propagation Graph}
\label{sec:propagation}

The first tool we present to the user is the Propagation Graph: 
a novel visualization which shows who broke the story on Twitter, and highlights influential and independent content creators.  The burstiness algorithm (described in section ~\ref{burstiness}) is used to identify the time when the story breaks, and the propagation graph shows the first hundred tweets in the breaking interval.  The Propagation Graph for the ``Plane in the Sea'' story can be seen in Figure~\ref{fig:propagation}.

A data point in the Propagation Graph represents a single tweet, and is plotted in several dimensions: the x-axis, which shows {\em time}; the y-axis, which shows the number of {\em retweets} received; and the size of a point, which represents the number of {\em followers} the tweeter has (scaled logarithmically).  Tweets written by {\em verified} accounts are marked by a bright blue border.  We claim that these are key elements in gauging the visibility of the tweet, as well as the degree of credibility other users will assign to the tweet and the amount of trust in the user as a source of information.  Since we are trying to track the flow of a story, time is a natural factor to observe.

But there are more dimensions that are depicted on the Propagation graph. Tweets with similar language 
(based on cosine similarity)  
have the same color, in an attempt to visualize {\em content independence}. 
More variation in color indicates there are likely multiple sources tweeting about the story: several different articles, or many individuals using different phrasing to talk about the same subject.
On the other hand, many tweets without variation in their wording may be a reason for suspicion.   
The tweet text might be the headline of an article with short commentary, a single tweet being copied and modified, or even a single person spamming the same information and varying the language very little as we have detected in several occasions.

The web interface allows users to view the tweets represented by points on the graph by hovering over or clicking on the points.  Studying the Propagation Graph (Fig.~\ref{fig:propagation}), we discover some facts about how the ``Plane in the Sea'' story developed.  Looking at the graph as a whole, we can see that the tweets, spread over only 20 minutes, are varied in content (many different colors) and number of retweets, and the users who have written them also vary in the number of followers they have. There is also a certain number of verified users, mostly news organizations.  


In this case study, the originating tweet appears as the first one in the graph.
It also happens to be the investigative tweet, reporting at 10:53 am EST that there is a plane in the sea near Telde in the Canary Islands, with a blurry picture of what appears to be an airplane in the water.  Information about the originator, the user who started the rumor by tweeting the picture of the airplane, is provided as well: @rafaleonortega's describes himself as a sports reporter, and has a moderate number of followers.  @rafaleonortega is not perhaps the most well-positioned source for breaking news, but his tweet has almost 600 retweets, so his message has been fairly well propagated, likely due to the accompanying comlelling image.

The next few tweets have similar messages, talking about a plane crashing in the sea, including a tweet at 11:01 am from @112canarias, a verified account tweeting about emergency information in the Canary Islands (112 is similar to 911 in the US).  This tweet confirms that a plane splashed down off the coast, though they do not know the number of passengers.  However, less than ten minutes later, at 11:09 am, @112canarias tweets again, now reporting that what was mistaken for a plane is actually a tug boat; at the same time, other verified accounts continue to report that a plane has crashed in the sea.  Two tweets from unverified accounts (@controladores at 11:13 am with 690 retweets, and @laprovincia\_es at 11:14 am with 239 retweets) also report that the plane crash is false, while more accounts continue to report about the crash.

\begin{figure*}[ht!]
\centering
\includegraphics[width=6in,keepaspectratio]{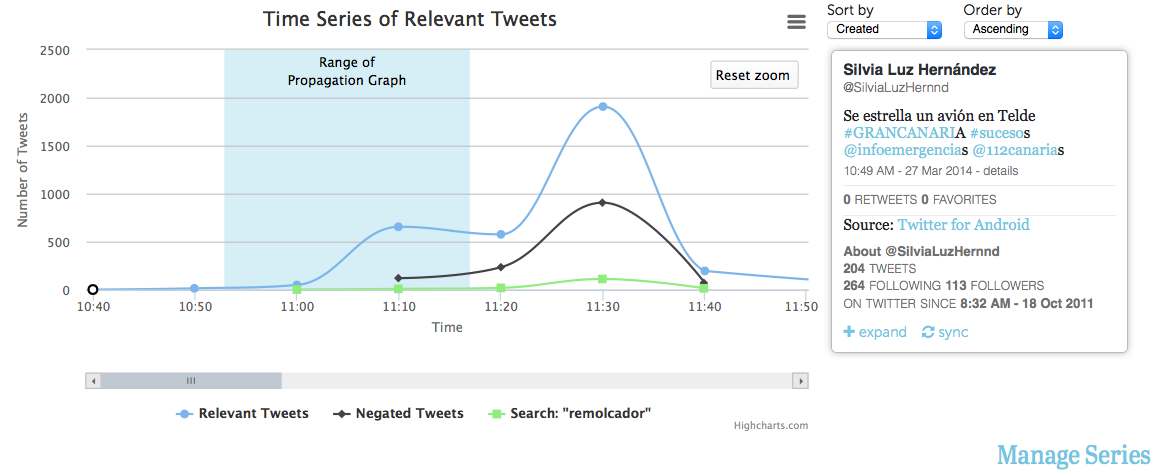}
\caption{The Timeline visualization from the ``Plane in the Sea'' story. Selecting a data point brings up a pane with all the tweets sent during this 10-minute interval. Three series are shown in this graph: all the relevant tweets, the negating tweets, and those the user chose to search for containing a particular keyword: remolcador (tug boat). It appears here that the negating tweets have succeeded in affecting the propagation of the rumor.}
\label{fig:time_series}
\end{figure*}

\subsubsection{Timeline Visualization} 
\label{timeline_section}
The Propagation visualization gives a detailed look into a specific interval of time: when the story broke; the Timeline visualization gives an overview of the whole story.  
An example is shown in Figure~\ref{fig:time_series}. 
The user can selectively browse the data in the timeline visualization without being overwhelmed by thousands of tweets.
Selecting a point on the graph will display the tweets written in that 10-minute interval to the right of the data point.  They can  be sorted in ascending or descending order by the number of retweets received, the time they were created or whether they are original tweets or retweets.  
The negation tweets (discussed in Section~\ref{sec:negation}) are also displayed as a series in this graph, to show when tweets denying the story began to spread.  



Although the propagation graph shows @rafaleonortega breaking the story, the time series shows a tweet written four minutes earlier by @SilviaLuzHernnd, another local journalist, claiming there was a plane crash, and mentioning two emergency information accounts, @112canarias and @Infoemergencias.  But while she has sent the first tweet in our dataset, we do not consider @SilviaLuzHernnd as the originator: her tweet was not retweeted, so the negligible visibility it received makes her an unlikely originator.  

According to the Timeline Graph, the story begins to pick up in popularity after 11:10 am, with over 500 tweets in a ten minute period.  The story peaks in popularity at 11:30 am, with almost 2,000 tweets in ten minutes.  The first mention of a tug boat can be seen at 11:09 am, preceding the negating information which claims the story of the crash is false.  Although the number of tweets negating the story never equals the number claiming it to be true, as the number of negating tweets increases, the number supporting it begins to decrease.

\subsubsection{Retweet and Co-Retweeted Networks}
\label{sec:rt}

The first two visualizations focus on the propagation of content over time.  {\sc TwitterTrails} also allows users to investigate the relationship of networks of users retweeting the story. The next two visualizations, the Retweet Network and the Co-Retweeted Network, help to answer questions about the main actors who were spreading information.

\begin{figure*}[ht!]
\centering
\includegraphics[width=6in,keepaspectratio]{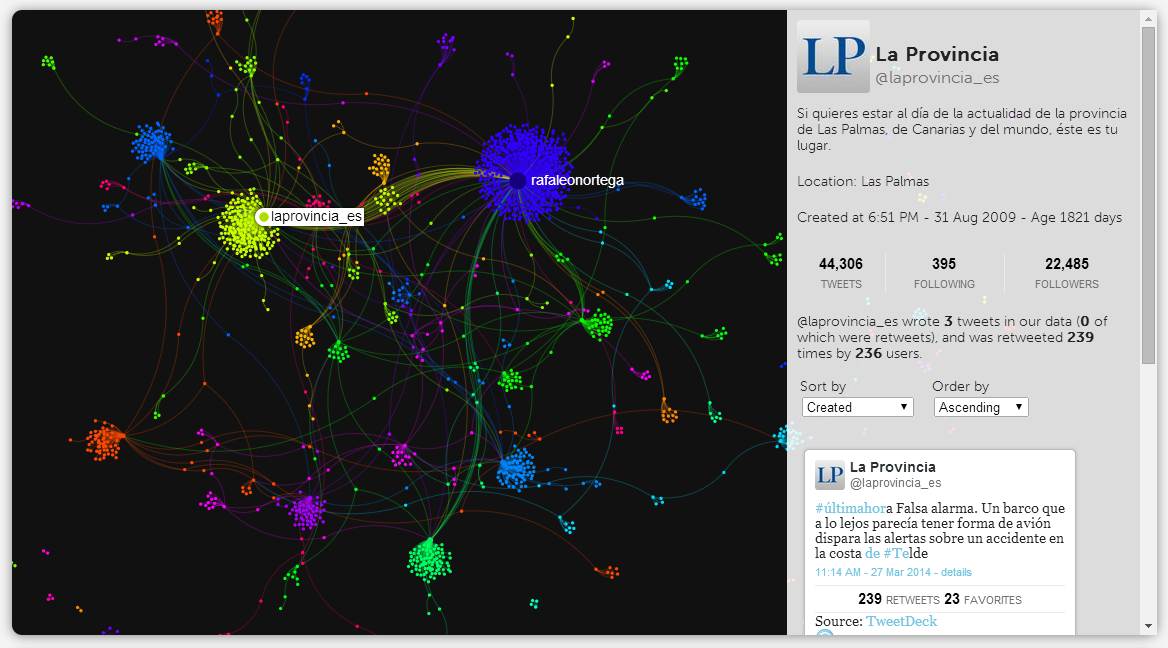}
\caption{The retweet visualization from the Plane in the Sea story.}
\label{fig:retweet}
\end{figure*}

\begin{figure}[ht!]
\centering
\includegraphics[width=3in,keepaspectratio]{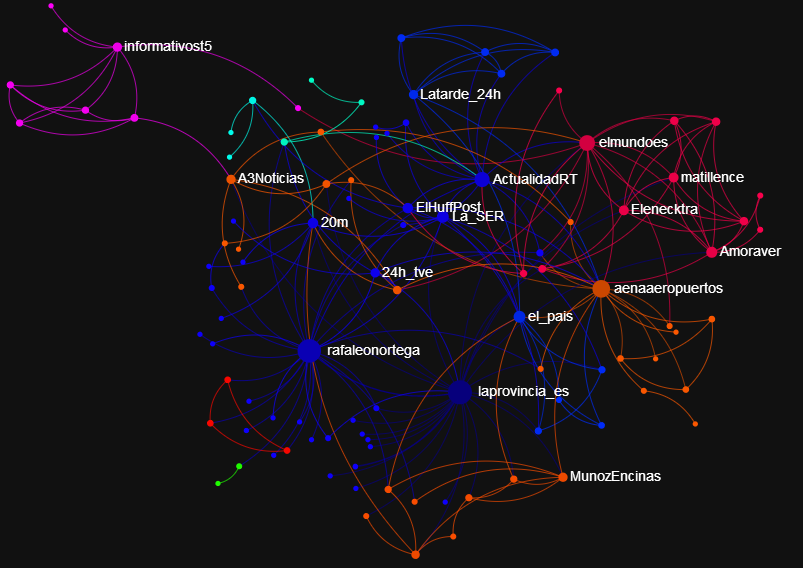}
\caption{The co-retweeted network from the Plane in the Sea story.}
\label{fig:corted}
\end{figure}

The two networks are created using Gephi~\cite{gephi}, where each node represents a user.  In the retweet network (Fig.~\ref{fig:retweet}), an edge between two nodes represents one user having retweeted the other.  The edges curve clockwise from the retweeting user to the retweeted user.  In the co-retweeted network on the other hand (Fig.~\ref{fig:corted}), edges represent two users having both been retweeted by a third user (see ~\cite{corted} for more discussion on the co-retweeted algorithm). Larger and darker nodes have a higher degree, and nodes are colored based on the modularity algorithm which groups nodes which are closely connected in the network ~\cite{corted}.


Figure~\ref{fig:retweet} shows the Retweet Network Visualization of the Plane in the Sea case study.  On the left is the retweet network graph, while on the right is an information panel which describes the node which the user has clicked on (in this case,  @laprovincia\_es): it shows information supplied by the user on their Twitter profile, as well as information about their tweets in the relevant tweet dataset.

The users who received the most retweets appear largest and most prominent in this graph.  @rafaleonortega appears as the most retweeted node in the graph, by 554 different users in the dataset.  @laprovincia\_es is also highly visible, with 236 users retweeting its tweets.  These two users were highlighted before in the Propagation visualization (Figure~\ref{fig:propagation}), where @rafaleonortega has spread the rumor of the plane crash, and @laprovincia\_es tweeted denying the rumor.  
Note that the clusters of accounts retweeting each of these two accounts are mostly not overlapping, indicating that each group has heard either the original story or its refutation.
However, there is a smaller group of accounts retweeting both @rafaleonortega and @laprovincia\_es. These are the users who propagated the initial false information and then propagated its correction. Clicking on these users on the web interface reveals that they have almost all retweeted first @rafaleonortega and then, minutes afterwards, retweeted @laprovincia\_es' refutation. This lends credibility to @laprovincia\_es' information: even though less users have retweeted him, his information that the crash was a false story was conclusive for many users.


The co-retweeted network shown in Figure~\ref{fig:corted} highlights the main actors from the retweet network, by connecting accounts based on mutual retweeting users.  That is, if User A and User B in the co-retweeted network and connected by an edge, it means at least one other user has retweeted User A and retweeted User B. @rafaleonortega and @laprovincia\_es are connected in the co-retweeted network because of the users who retweeted both of them.  Connections indicate related content: in this case the relationship is that the content created by @laprovincia\_es is a response and contradiction of information from @rafaleonortega.

\subsubsection{Interaction Between the Visualizations}
Each of these visualizations can be considered independently of the others. However, one of the goals of {\sc TwitterTrails}  is to create a complete picture of the investigative story.  One step in doing this is connecting the visualizations. Interacting with any of the visualizations will highlight corresponding information in the other visualizations, to help the user utilize all of the tools to trace a specific user or tweet.

\subsubsection{Tweeted Link Bibliography}
Although {\sc TwitterTrails} limits its search of relevant information to Twitter, many users will insert urls into their tweets as a way of citing the blogs and news articles that they got their information from.  The Tweeted Link Bibliography counts the most cited links, as well as how many users tweet the links, and provides an interface for exploring the tweets containing that link.  

In the ``Plane in the Sea'' story, the most linked pages are all news posts about how there was no plane crash and the rumor was a false alarm.

\subsubsection{Summary of an Investigation}

Finally, {\sc TwitterTrails} produces a summary in the form of a report that refers to the main findings of the algorithms and the values of the metrics described in section \ref{algorithms}. The text of the summary (Figure~\ref{fig:summary}) tries to answer the questions we originally posed in the Introduction, but also give a sense on whether the reaction of the audience of the story on Twitter have any doubts about the truthfulness of the rumor. In most cases, this summary is enough to get a good idea of the characteristics of the rumor. 

In addition, {\sc TwitterTrails} produces two views of all the stories we have investigated so far (more than 100 at the time of the writing). In a condensed view (Fig.~\ref{fig:twittertrails-com}, top) the investigated tweet is shown along with the calculation of the overall visibility of the story and the skepticism of the audience. In the full-view (Fig.~\ref{fig:twittertrails-com}, bottom) the user can explore all the findings of the system.

\section{Algorithms and Metrics} 
\label{algorithms}

There are two main algorithms powering the {\sc TwitterTrails}  tool: first, the algorithm which selects the relevant tweets that go into building the visualizations, controlled by parameters set by the user; and second, an algorithm to detect when the story first broke on Twitter.  Of interest may also be the co-retweeted algorithm which we use to build a network graph (see Section ~\ref{sec:rt}), which is presented in \cite{corted}.

\subsection{Tweet Relevancy Algorithm}
\label{relevancy}
The first step to creating a story is to collect the relevant data which forms the story.   The stories {\sc TwitterTrails} creates are made up of tweets which are relevant to the user's investigation.  
Finding so called ``relevant'' data necessitates considering the balance as to how relevancy is defined. 
It is important that the definition of a "relevant" piece of data should be broad enough to capture interesting and important facets of the story, but limited so that the dataset is manageable for human consumption.  {\sc TwitterTrails}  attempts this task by employing a user-controlled data filter based on keywords from the investigative tweet and the user's own knowledge.

The user may control the collection of data by selecting words or phrases (which we refer to as "keywords") from the text of the tweet they are investigating, or by manually inputing keywords absent from the text. 
{\sc TwitterTrails} then takes the keywords chosen by the user to collect tweets via Twitter's Search API
with each keyword being used as a search query.  The number of tweets collected can be controlled by the user as well.  The Search API returns recent tweets containing words in the query in reverse chronological order (newer tweets are returned first), but is limited to tweets written or retweeted in the last 6-9 days.
Because of this limitation, {\sc TwitterTrails} is best suited to investigating recent and breaking stories.

The keywords the user selects are used by the Search API to collect tweets (the number of tweets can also be adjusted by the user, up to 18,000 tweets/keyword).  
From the data retrieved by the Search API, {\sc TwitterTrails} automatically calculates which are the relevant tweets based on other inputs the user controls.  The user can require a relevant tweet to contain some subset of keywords.  
The user defines keywords as being either ``required'', ``optional'' or ``excluded.''  Required keywords must appear in a tweet for it to be relevant, and if the user choses multiple required keywords, she can select whether all or at least one must appear in a relevant tweet.  
The optional keywords are controlled by a threshold (also defined by the user): a relevant tweet must contain at least the number of optional keywords set by the threshold.  The threshold can be set to 0, in which case the optional keywords have no effect on the relevancy algorithm.  Finally, any tweet containing any of the excluded keywords will not be considered relevant (even though it may have the optional or required keywords).  
In addition to the keyword-based inputs, the user can define a time period to limit relevant tweets to. Any tweets outside of the time period will not be considered relevant.  

In many investigations the initial set of search keywords must be modified to define relevant tweets. {\sc TwitterTrails}  gives the user the option to redefine this set as many times as she wants until she has discovered the best set of relevant tweets to study.

\subsection{Burstiness Algorithm} 
\label{burstiness}
One of the purposes of {\sc TwitterTrails} is to give users tools for investigating the origins of a story on Twitter: who broke the story and when.  {\sc TwitterTrails} automatically identifies the moment a story breaks on Twitter by computing the time interval in which relevant activity in the story increases significantly.  

To identify this moment, we look for above-average bursts in activity over $N$ 10-minute intervals.  
 {\em Activity} $A_n$ during interval $t_n$ to $t_{n+1}$ is defined as the total number of retweets received by tweets written in the $n$-th interval (includes retweets received outside of the interval).  This means that an interval $A_1$, in which 100 tweets were written with only one retweet each, is considered less active than an interval $A_2$, which contains one tweet with 1000 retweets.  

 We measure the $burstiness_n$ of the $n$-th interval by calculating the cosine of the angle between this and the preceeding $(n-1)$-st interval:

 \[burstiness_{n} = 1 - \frac{t_{n} - t_{n-1}}{\sqrt{(t_{n} - t_{n-1})^{2} + (rt_{n} - rt_{n-1})^{2}}}\]

 where \(rt_{n}\) is sum of the retweet count of tweets written during interval \(n\). This only counts original tweets; retweets do not contribute to \(rt_{n}\).  The result is a value from 0 to 1, where higher values indicate a stronger burst.  One special case is where \(n\) equals 0.  In that case we replace \(rt_{n-1}\) with the average value of \(rt\) across all $N$ intervals.

We consider activity as the sum of the retweet counts because this correlates to how much attention a tweet received and how many people have been exposed to it.  Our algorithm does not identify the first tweet to post relevant information as breaking the story, but rather the first tweet to make a sizeable impact; if the first tweet received very little attention, then we don't consider it to have broken the story.

Once we have identified the initial burst in the data, we visualize these tweets in the ``Propagation Graph'' described in Section ~\ref{sec:propagation}, to allow the user to study them in more detail and answer questions about how the story originated and how information propagated when the story broke.

\subsection{Negation Classification}
\label{sec:negation}
Another question {\sc TwitterTrails} allows the user to investigate is whether there is doubt about the veracity of a story circulating, and how  has it spread.  This is done by calculating ``negation" tweets\footnote{Refutation is a better term, but we use negation to hint at the method we employ to compute it.}: those tweets which express doubt or claim information is false or a hoax. We are not aware of a reliable method to compute sentiment of a tweet that classifies negation. 


State of the art techniques in sentiment analysis \cite{kiritchenko2014sentiment} use negation markers to establish contexts in which the polarity of sentiment words might be reversed.  
The list of negation markers is limited, but their interaction with different sentiment words complicates reliable calculation of sentiment scores. Due to the size limitation of tweets, negating a story often is done through a simple vocabulary of words, such as fake, untrue, hoax, debunked, etc. For the goals of our system, we find that it's not necessary to do a sophisticated sentiment analysis, since the system is only interested in meta information about the story (its veracity), as opposed to the sentiment expressed in the claims. 


Currently we employ a simple classification algorithm which identifies tweets with a small set of keywords that indicate negation, doubt, or denial, such as ``hoax,'' ``fake,'' and ``untrue.''  In addition, we provide the user the ability to add or remove from these keywords on a story by story basis; for example, when tweets are in another language or a user finds terms specific to the negation of a story. We note that while we have provided this feature, we very rarely make use of it in the data we have collected so far. 

The negation tweets can then be visualized in the Timeline graph (described in Section~\ref{timeline_section}).  They are also used to calculate a Skepticism score, based on how much negation tweets are propagating in the story compared to non-negation tweets (in Sect.~\ref{sec:metrics}).

Our negation algorithm is arguably not very sophisticated, as it will miss expressions of sarcasm, irony and slang. Yet, our experience shows that it has reasonably good performance. We evaluated at random a couple stories and found it to have accuracy around 80\%. There is certainly room for improvement and a better classification remains as an open problem. However, the success of this simple classification may be enabled by the limited number of characters a tweet offers for debunking, so English speakers resort in using a limited dictionary.

\subsection{Propagation Metrics}
\label{sec:metrics}
We have introduced a couple of metrics to measure the propagation of a story on Twitter: The {\em propagation score} and the {\em skepticism score}. Inspired by the relevant theory of Library Science, we consider a tweet as a ``publication'' and its verbatim retweets as its citations, evidence of its propagation in the network (e.g., \cite{corted}). The h-index\footnote{Wikipedia entry for h-index http://en.wikipedia.org/wiki/H-index} of a collection of $N$ publications 
is defined to be $h$ when there are $h$ publications in the collection that each have at least $h$ citations and the remaining $(N-h)$ publications have less than $h$ citations each.

In a similar fashion this approach allows us to measure the {\em propagation score} of a collection of tweets (e.g., those retrieved as relevant to a story) as its h-index. We say that a story has h-index $h$ if there are $h$ tweets that have received at least $h$ retweets. We then define the {\em propagation level} of a story on a discrete logarithmic scale of h-indices: A story with h-index of at most 16, 32, 64, 128 is said to have propagation level of {\em insignificant, low, moderate, high}. There are very few stories that achieve h-index above 128, and they are said to have propagation level {\em extensive}. At this moment, the highest score is 444 for the report of Robin Williams death.

Note that propagation level can be defined on any meaningful collection of tweets. In particular, we can defined it on the subset of negating tweets and non-negating tweets as found by our negation classification (Section~\ref{sec:negation}). We thus define as the {\em skepticism level} of a story the ratio of the negation h-index over the non-negation h-index of the story. Coincidentally, at this moment the most objected-to story is 3.2, about the hoax announcing the death of Judd Nelson's death.

\begin{figure*}[ht!]
\centering
\includegraphics[width=6.0in,keepaspectratio]{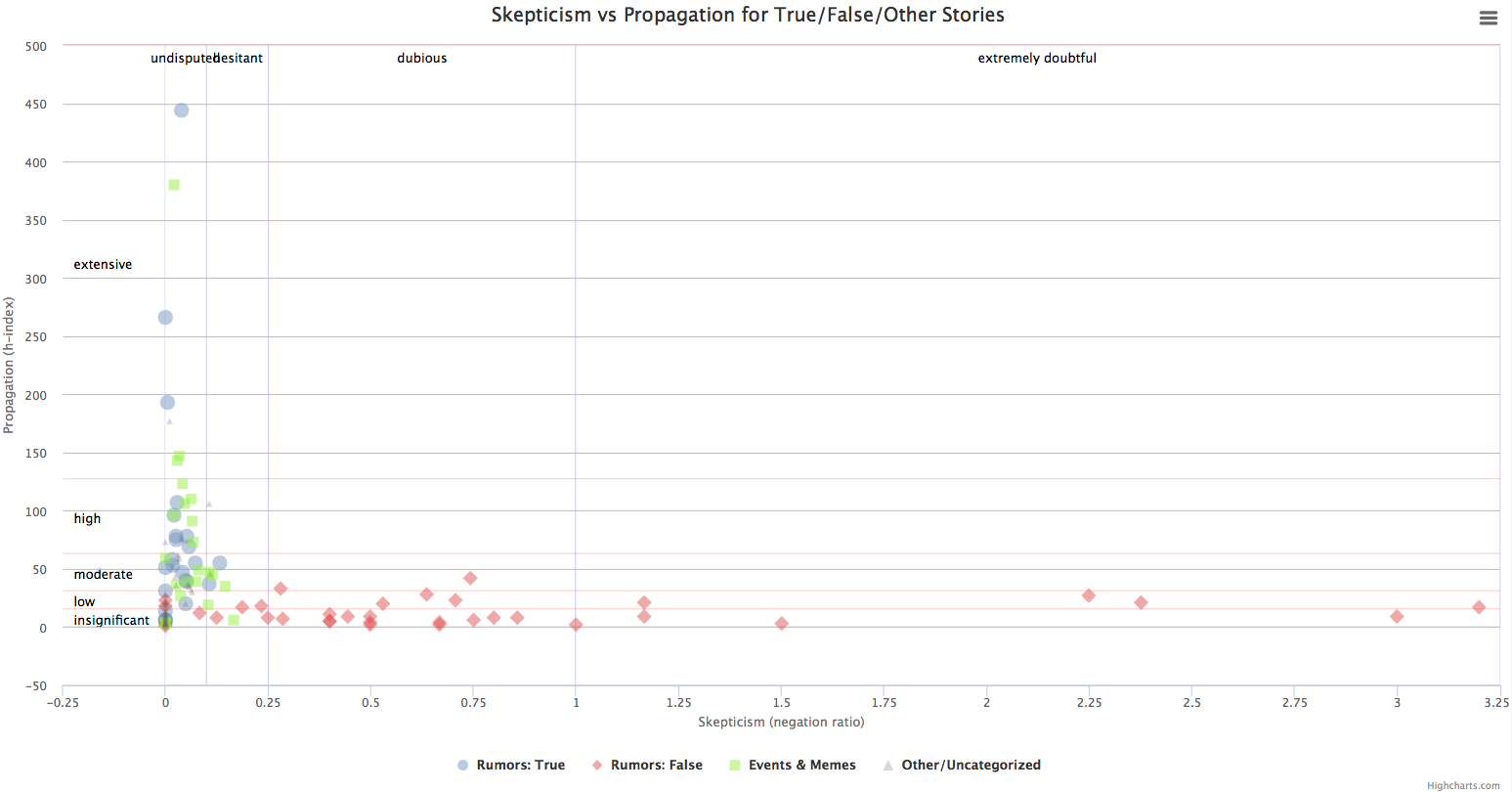}
\caption{False rumors on Twitter exhibit low propagation, high skepticism scores (see Section~\ref{sec:discussion}.}
\label{fig:propagationVSskepticism}
\end{figure*}

Plotting the propagation level versus the negation level for our collection of (currently over 100) stories  we see an L-shape of data points (Fig.~\ref{fig:propagationVSskepticism}). In particular we observe that stories do not score high in both propagation and negation levels. We discuss this observation in the  section~\ref{sec:discussion}.

\section{True and False Rumors}
\label{sec:discussion}
We are currently using {\sc TwitterTrails} to collect and analyze an increasing number of stories as they come to our attention. In addition, we are collecting stories that have gotten some attention in the news, such as those related to political events, crimes, or defined by hashtags. Moreover, we are collecting stories reported at the Emergent site \footnote{Emergent, a project by Craig Silverman  http://emergent.info} and Washington Post's ``The Intersect'' page. \footnote{``The Intersect'', Washington Post blog by Caitlin Dewey http://wapo.st/1wINhTj}
There is certainly a bias in our collection of stories towards those false ones that receive the  interest from journalists and the media (one has to try very hard to find false claims on any day, compared to the volume of information propagated on Twitter), but we believe that this bias strengthens our results. 

We divide the stories in four categories (See Fig.~\ref{fig:propagationVSskepticism}):
\begin{itemize}
\item
{\bf Rumors: True/False.}
They stem from tweets stating verifiable facts and can be determined, at some point, to be either True or False (two categories). As example of a false rumor is our ``Plane in the sea'' story, which was believed for a while to be true by the official emergency site in the Canary Islands. As an example of a true rumor is that of a 9 year old girl killed her gun instructor.

\item
{\bf Events and Memes.}
Those related to Events and Memes, that is, data collected by simply searching for a hashtag such as \#VASEN (related to the recent senatorial elections in Virginia, USA) and \#alexfromtarget (related to a meme regarding a photograph of a celebrity look-alike boy). 

\item
{\bf Other/Uncategorized.}
This is a default category for searches that do not fall in the two categories above. They may include opinion (e.g., a certain celebrity is the hottest man alive) or are simply uncategorized so far.  
\end{itemize}

\subsection{Is it true or false? Ask the crowd!}
We now focus on the verifiable True/False rumors. In establishing the ground truth we utilize reporter's work, such as the findings by Snopes\footnote{Snopes, http://www.snopes.com}, Emergent and The Intersect, or do our own investigation\footnote{TwitterTrails Blog, https://blogs.wellesley.edu/twittertrails/}. Our ongoing research indicates a very interesting trend, which we conjecture to be true: 

\begin{quote}
{\bf Conjecture:} {\em On Twitter, claims that receive higher skepticism and lower propagation scores are more likely to be false. On the other hand, claims that receive lower skepticism and higher propagation scores are more likely to be true.}
\end{quote}

Intuitively, this conjecture can be explained as an example of the power of crowdsourcing. Since the ancient times philosophers have argued that people will not willing do bad unless they are guided by irrational impulses, such as anger, fear, confusion or hatred\footnote{``The impossibility of Acrasia'', Stanford Encyclopedia of Philisophy  http://plato.stanford.edu/entries/plato-ethics-shorter/\#9
}. Therefore, the more people see some false information, the more likely it is that they will either raise an objection or simply decide not to repeat it further. 

We make the conjecture specific for Twitter because it may not hold for every social network. In particular, we rely on the user interface for promoting an objection to the same level as the false claim. Twitter's interface does that; both the claim and its negation will get the same amount of real estate in the a user's Twitter client. On the other hand, this is not true for Facebook, where a claim gets much greater exposure than a comment, 
while a comment may be hidden quickly due to follow up comments. So, on Facebook most people may miss an objection to a claim. This may explain the result by \cite{rumor-cascades} who find that false claims live for a long time, even those that are verifiable by a quick search on Snopes. By contrast, very few Snopes-included claims appear on Twitter.

\section{Related Work}
\label{related}

Big data necessitates sense making tools, especially in the form of interactive visualizations, to allow humans to process and interpret the data. Pirolli et al. studies how information should be organized for intelligence analysis, and to this end introduces a "sense-making loop" describing the process in which a tool like {\sc TwitterTrails}  can help its user analyse information \cite{pirolli}.  They postulate that visualizations should be used as a sort of external memory for a user, to improve a user's memory and processing capabilities.  {\sc TwitterTrails}  follows a bottom up data process similar to the one described in the paper, in which data is gathered and refined with human input, and then creates visualizations to help the user filter and consume the data in a meaningful way, to be able to formulate theories about the data. 

Similar to {\sc TwitterTrails}  are tools which focus on timelines to visualize the spread and propagation of a story or real time event, often focusing on bursts or peaks of data to assist in summarization of the data.  Narratives tracks the frequency of terms in blog data to track the evolution of news stories \cite{fisher}.  Like in {\sc TwitterTrails}, Chieu et al. use burstiness algorithm to automatically detect peaks in the data, and use these to extract and summarize events, and then rank them based on interest and importance \cite{chieu}.  Although {\sc TwitterTrails}  performs a similar task in finding the first peak in the data, its end result is to provide the propagation visualization to the user, in order to allow them to analyse and theorize about the origin of their story.

Some of these tools focus on highlighting keywords and phrases in the data: ThemeRiver uses a timeline to map the prominence of topical keywords overtime, to find temporal patterns quickly and easily \cite{havre}. A similar meme tracking tool is developed by Leskovec et al., mapping the rise and fall of memes in the blogosphere and news media \cite{leskovec}.  TimeMines creates “overview timelines” by extracting nouns and named entities and charting the frequency of these features over time \cite{swan}.  

So far, these tools take data from news media, blogs and internet searches.  
{\sc TwitterTrails}  focuses on the spread of data on social media, generated by both official news media and individuals reporting and spreading stories.  
Social media websites like Twitter generate massive amounts of data every day. 
Twitter’s rise to prominence in both daily and professional life has led to the creation of many tools to make sense of its data \cite{marcus, bernstein,voxcivitas}.  Eddi mines a single user’s stream, which can be overwhelmed by hundreds of tweets daily, and utilizes topical analysis for users browse their stream \cite{bernstein}.  TwitInfo is a visual interface to assist users in summarizing event specific data, such as a sports match \cite{marcus}.  
It aggregates topical data from Twitter’s Streaming API and automatically identifies and labels peaks in the data.  It uses a timeline interface similar to {\sc TwitterTrails}  to allow users to browse through data, but adds sentiment and geolocation to give more information about the data. 
Vox Civitas, a graphical tool created by Diakopoulos et al., has a similar motivation as {\sc TwitterTrails} : to assist users, specifically journalists, extract interesting and meaningful information from social media streams \cite{voxcivitas}.  
Their tool mines query specific data from media events on Twitter, and visualizes both sentiment and topics over time.  When journalists evaluated Vox Civitas, one reaction was that they would use it to track sources of data.  {\sc TwitterTrails}  focuses on this goal, using a similar time series graph to allow users to navigate through Twitter data. 
In the same vein, Videolyzer highlights claims in videos, and positive and negative reactions to these claims, to help bloggers and journalists analyze the quality of user created videos \cite{videolyzer}.  

Monitoring and evaluating the propagation of a rumor has recently gotten a lot of attention. A new web service, {\tt emergent.info} developed by journalist Craig Silverman is using journalists to evaluate online claims\footnote{``Why Rumors Outrace the Truth Online'' by Brendan Nyhan, Sept. 29, 2014 
http://nyti.ms/1pFXaAq}
and deem them as True/False/Unverified. They track the number of shares a rumor has on Facebook, Twitter and Google+ and report the numbers along with links to articles that supported and countered the rumor.
Another application is RumorLens \cite{AJR-rumorlens} by Paul Resnick et al. It analyzes the spread of rumors on Twitter and prompts user feedback to classify results as propagating, debunking or unrelated to the original rumor. It will then use a text classifier to garner more widespread results. 

Rumor Cascades on Facebook have also been studied by a Facebook team \cite{rumor-cascades}. They are focusing on tracking the way that rumors propagate on Facebook, mainly those that have been verified in snopes.com. Unlike what we observe on Twitter, they find that rumors do not easily die on facebook but they may emerge long after they started.

The propagation of false rumors related to a Chilean earthquake was analyzed by \cite{Chato-info-credibility} where they found that there are measurable differences in the way messages propagate, that can be used to classify them automatically as credible or not credible. Further, \cite{TweetCred} reports on the development of a plug-in that, based on fixed characteristics of a tweet it is able to quickly evaluate credibility scores for any tweet without reference to its type (rumor, event, opinion, etc.)

One of the earliest systems that focused on studying patterns of information propagation in online social networks like Twitter is Truthy 
\cite{ratkiewicz2011truthy}. Truthy is based on the concept of memes that spread in the network. Such memes are detected and followed over time to capture their diffusion patterns. Truthy is a more general-purpose system than the ones we mentioned previously in this section, which, despite its name, it doesn't provide explicit assessment of the veracity of the tracked memes. However, through visualizations of propagation patterns and other metrics (e.g., sentiment analysis), Truthy can enable a user to come to a certain conclusion on her own.

\section{Conclusion and Future Work}
\label{conclusion}
{\sc TwitterTrails} was designed and implemented with the goal to provide a vital service to users who want to engage with Twitter as a source of reliable information, either for their own consumption, or as a source for journalism, both professional and amateur.  

{\sc TwitterTrails} makes it easy to investigate a suspicious story. By inputting a single tweet into the system, and selecting keywords relevant to the story being investigated, the system will gather a dataset of tweets through which the user can trace the story origin. The Tweet Propagation visualization 
focuses on the moment the story first broke on Twitter, while the Timeline Visualization on how it spread.  Both of these allow users to meaningfully and easily sort through hundreds to thousands of tweets, and highlight both tweets and periods of time that are most interesting to the story.  The two network visualizations, a Retweet and a Co-Retweeted network, allow users to study accounts on Twitter who were both influential propagators of information, and sources who other users put trust in.  

Our system leads us to conjecture that true and false rumors have different footprints in terms of how they propagate and invoke skepticism by their audience. False rumors are more likely to be negated if exposed to a larger audience.

The most pressing area of future study for {\sc TwitterTrails}  is to design and implement a user evaluation of the tool, to further improve its functionality and usefulness. We plan to draw inspiration from Kang et. al ~\cite{kang}, who outline a method for in depth evaluation of visual analytics systems.  They log and analyse user activity not only using the system to be evaluated, but also with more low tech approaches to solving the same problem.  
We also plan on evaluating and improving our algorithms to detect when a story breaks on Twitter, and filtering for relevant tweets.  We also hope to pursue more methods of customizing {\sc TwitterTrails}  for users in ways specific to the story they are investigating.  This may include creating more visualizations, which the users can pick and chose from as is appropriate to their investigation, and creating more meaningful ways in which these visualizations can interact with each other.

We close this section with a note and invitation that our system is open, our collection is not based in proprietary data, our methods simple and easily implementable, and interested researchers can replicate and verify our work.

\section{Acknowledgements}
This research was partially supported by NSF grant CNS-1117693 and by the Wellesley Science Trustees Fund. The authors would like to thank the students who have worked in the implementation of the system, especially Laura Zeng, Lindsey Tang, Susan Tang, Megan O'Keefe, Christina Pollalis.


\end{document}